\documentclass[journal,final,a4paper]{IET-Conf-Paper}
\usepackage{IEEEtrantools}

\usepackage{caption}
\usepackage{subcaption}

\usepackage{xspace}
\newcommand{\TSO}{\textsc{TSO}\xspace}
\newcommand{\DSO}{\textsc{DSO}\xspace}
\newcommand{\TSOs}{\textsc{TSO}s\xspace}
\newcommand{\DSOs}{\textsc{DSO}s\xspace}
\newcommand{\OPF}{\textsc{OPF}\xspace}
\newcommand{\ADN}{\textsc{ADN}\xspace}
\newcommand{\ADNs}{\textsc{ADN}s\xspace}
\newcommand{\TSODSO}{\textsc{TSO-DSO}\xspace}
\newcommand{\SOCP}{\textsc{SOCP}\xspace}
\newcommand{\acopf}{\textsc{\mbox{AC--OPF}}\xspace}
\newcommand{\distflow}{\textsc{\mbox{DistFlow}}\xspace}
\newcommand{\distflowSOCP}{\textsc{\mbox{DistFlow--SOCP}}\xspace}
\newcommand{\lindistflow}{\textsc{\mbox{LinDistFlow}}\xspace}

\usepackage{breqn}
\usepackage{amsmath}
\usepackage{graphicx}
\renewcommand{\figurename}{Figure}
\usepackage{algpseudocode}
\usepackage{algorithm}
\usepackage{multirow}
\usepackage{siunitx}
\usepackage{xcolor}

\usepackage{float}
\floatstyle{ruled}
\newfloat{model}{thp}{lop}
\floatname{model}{\textsc{Model}}

\usepackage[intoc]{nomencl}
\makenomenclature

\usepackage{etoolbox}
\renewcommand\nomgroup[1]{%
  \item[\bfseries
  \ifstrequal{#1}{I}{Indexes}{%
  \ifstrequal{#1}{S}{Sets}{%
  \ifstrequal{#1}{V}{Variables}{%
    \ifstrequal{#1}{A}{Acronyms}{%
  \ifstrequal{#1}{P}{Parameters}{}}}}}%
]}

\begin{document}

\title{Coordination Between TSOs and DSOs: \\ Flexibility Domain Identification}
\author{Liliia Ageeva\ad{1}, Maryam Majidi\ad{2}, David Pozo\ad{1}\corr\ }

\address{
\add{1}{Center for Energy Science and Technology, Skolkovo Institute of Science and Technology (Skoltech), \\Moscow, Russia}
\add{2}{Department of Energy Management and Operation of Electrical Networks, University of Kassel, \\Kassel, Germany}
\email{d.pozo@skoltech.ru}}


\begin{abstract}

The enormous technological potential accumulated over the past two decades would make possible to change the operating principles of power systems entirely. The consequent technological evolution is not only affecting the structure of the electricity markets, but also the interactions between Transmission System Operators (TSOs) and Distribution System Operators (DSOs). New practical solutions are needed to improve the coordination between the grid operators at the national, TSOs, and local level, DSOs.
In this paper, we define the flexibility range of coordination between TSOs and DSOs. 
By doing so, we propose an algorithm based on epsilon-constrained methods by means of mathematical programming and power systems principles. 
We evaluate and compare different classical optimal power flow formulations (\acopf, \distflow, \distflowSOCP, and \lindistflow) for building the flexible \TSODSO flexible domain.  The presented approaches in this paper are analyzed in an IEEE 33-bus test radial distribution system. We show that for this particular problem, the \distflowSOCP has the worst accuracy, despite the popularity among the academic community of convex relaxation approaches.

\end{abstract}

\maketitle

\section{Introduction}

The current power energy industry is on the inception of, possible, the largest transformation
in its history. The enormous technological potential accumulated over the past two decades 
would make possible to change the operating principles of power systems entirely. The consequent
technological evolution is not only affecting the structure of the electricity markets, but also the
interactions between transmission system operator (\TSOs) and distribution system operator (\DSOs). A new generation of distribution networks are evolving to have an active role in the control and management of every
participant connected to it so, power is not anymore unidirectional between distribution and transmission grids. New practical solutions are needed to improve the coordination between the grid operators
at the national, \TSOs, and local level, \DSOs in a reliable, secure, and economic fashion. Solving this problem is vital for the future power system since the number of uncertainties
in the grid is rising as a result of widespread distributed  energy generation resources, mainly renewable. 

One of the crucial points of studying active distribution networks (\ADNs) is the flexibility that \ADNs can provide to the main grid, i.e., at the transmission level.
This can unpin capabilities of providing new services to \TSOs from \DSOs, like ancillary services.
The efficient coordination between \TSOs and \DSOs could bring some advantages such as congestion management, system balancing,  power quality control, enhance real-time control and supervision, and grid infrastructure updates shifting among others.

\subsection{State-of-the-Art} 
Literature related to the \TSODSO interaction is relatively new. The paper \cite{barenfanger2016classifying} proposes a classification scheme, which is called a taxonomy,  for the different types of flexibility, in both research and industrial projects, that are used in electric grids. The study in \cite{villar2018flexibility} reviews flexibility products and markets designs and implementations to support the power systems operation by considering the renewable generation and distributed energy resources increases. 
A modeling framework is presented in \cite{gonzalez2018determination} to approximate the flexibility of an ADN by the regulation of the power flow over the \TSODSO interface to provide ancillary services. Correspondingly, the effect of time-invariant influencing factors on the flexibility of the \ADN  is discussed. In \cite{contreras2018improved}, the aggregated flexibility of distribution grids without needing to release sensitive grid data is improved using linear optimization. The proposed model is validated using two real radial MV-level distribution grids in Germany. A robust distributed generation investment planning, which considers the uncertainties associated with the intermittent renewable generation and variable electricity demand, is proposed at \cite{ji2018robust} to minimizes the net present value of total costs. \cite{elliott2018sharing} validates an integrated communication and optimization framework for performing the coordination of a \TSO congestion relief with a \DSO objectives. \cite{ageeva2019analysis} proposes a methodology toward the calculation of the \ADNs flexibility based on the feasibility region at the \TSODSO interface by employing a Monte Carlo sampling approach. The paper in \cite{silva2018estimating} presents a methodology based on the solution of a set of optimization problems that approximate the flexibility ranges at \TSODSO margin while considering the technical limits and a maximum cost that the customer is willing to pay. The flexibility for \TSODSO plus Retailer coordination in Britain is evaluated in \cite{pastor2018evaluation}.

\subsection{Paper Approach and Contributions}
This paper defines a flexible region as all the possible values of active and reactive power at the interface \TSODSO, i.e., substation/s, such as a \ADN can operate without violating any technical limits within the \ADN. 
Thus, the principal objective of this work is to identify the boundaries of an active-reactive power at the interface \TSODSO. For doing so, we employ mathematical programming for finding optimal operating points where active and reactive power at the \TSODSO interface are cross-examined to their extreme values while keeping feasible the distribution grid operation. 
We propose an epsilon-constrained optimization method that, contrary to  Monte-Carlo-based methods, does not require a large number of simulations.
Besides, we argue that Monte-Carlo-based methods are not appropriate for feasible domain construction. The feasible domain should be built on the basis of the (single) optimal operating dispatch point by finding the largest capacity that flexible resources and distribution grid can provide by ``stressing''  them but not formed by different loading conditions that imply different dispatch.

The main contributions of this work are in two main directions. 
\begin{itemize}
    \item \textit{On the methodology}. We propose a methodology based on the epsilon-constraint method, adopted from multi-objective optimization, to construct the boundaries of the \TSODSO feasible regions that efficiently generate the feasible region with a very small number of simulations. Opposite to Monte Carlo methods, the number of simulations needed for feasible region construction does not depend on the control and uncertain parameters of the \ADNs; %
    \item \textit{On the analysis}  The above contribution is supported by an optimal power flow (\OPF). There are several \OPF models the literature. We identified four of the most common \OPF approaches for the optimal operation of \ADNs, so called, (i) alternating current optimal power flow, \acopf, (ii) an ad-hoc reformulation for the \acopf for distribution grids, \distflow, (iii) a convexified version of the \distflow model based on second-order cone programming, \distflowSOCP, and (iv) a linearized version of \distflow, \lindistflow. The epsilon-constrained method is tested in the IEEE 33-bus distribution network and compared for each \OPF model.
\end{itemize}

\subsection{Paper Organization} 
This paper is structured as follows. In  Section \ref{section2}, different \OPF methods in distribution grids are presented for a single-period case. 
Section \ref{section4} presents the epsilon constraint method as the main methodology to define \TSODSO  feasibility region. Section \ref{section5} presents the case study. Finally, conclusion is given in Section \ref{section6}.

\section{Optimal Power Flow in Distribution Grids} 
\label{section2}
The purpose of utilizing the \OPF is determining the optimal operating point for an electric power system relative to desired objectives, such as minimizing generation cost and losses. Along with that, the solution must fulfill the constraints that model the power flow physics and enforce technical limits \cite{wood2013power}. 
The general form of the cost-prioritized \OPF objective function is defined in \eqref{eq.obFunACOPF}. 
The first term represents the cost/benefits from the power requested/injected from/to the main grid. The second term represents the generation cost within the \DSO.
\begin{align}
 \min \: \left(  c^{\text{se}} p^{\text{se}}  + \sum_{i \in \mathcal{N}  \backslash \{1\}} c_i p^{{\text{G}}}_i  \right)  
\label{eq.obFunACOPF} 
\end{align}

 The cost term at the \TSODSO interface, $c^{se}$, is assumed to be symmetric in here. Thus, power withdraw from the substation has the same cost that the payment received for the power injected to the substation. The term $p^{se}$ takes positive value when power flows from \TSO to \DSO and negative when it flows from \DSO to \TSO. As customary in literature, we have assigned the bus number one to the substation. Also, we have modeled a single substation without loss of generality.  Various versions of the \OPF problem are introduced in the next subsections.

\subsection{\acopf Formulation}
The \acopf is the most common representation of the full AC power flow equations and operational limits associated with the power grid. It is based on the nodal power flows in a electric network.
The \acopf is a non-linear and non-convex optimization problem. Thus, no global optimum solution is guaranteed for this problem  \cite{molzahn2019survey}. 

The nodal balance equations for the active and reactive power are formulated in \eqref{eq2} and \eqref{eq3}, respectively.
\begin{IEEEeqnarray}{l"l}
  p_i^{\text{G}} - p_i^{\text{D}} - \sum_{j:(i,j) \in \mathcal{L}} p_{ij} = 0, & \forall i \in \mathcal{N} \label{eq2} \\
  q_i^{\text{G}} - q_i^{\text{D}} + q_i^C - \sum_{j:(i,j) \in \mathcal{L}} q_{ij} = 0, & \forall i \in \mathcal{N} \label{eq3} 
\end{IEEEeqnarray}

We have assumed a single generator and single load/demand at each node $i$, for the sake of simplicity.

The branch active and reactive power flow, $p_{ij}$/$q_{ij}$, are given  by equations \eqref{eq4} and \eqref{eq.5}, respectively.
\begin{IEEEeqnarray}{l'l}
p_{ij} =&  v_i^2 Y_{Lij} \cos(\theta_{Lij}) 
- v_i v_j Y_{Lij}\cos(\delta_i - \delta_j - \theta_{Lij})  \nonumber  \\
&+ \frac{1}{2}v_i^2 Y_{Sij} \cos(\theta_{Sij}), 
\qquad \quad  \qquad  \forall (i,j) \in \mathcal{L}
\label{eq4} \\
q_{ij} =&  - v_i^2 Y_{Lij} \sin(\theta_{Lij}) 
- v_i v_j Y_{Lij} \sin(\delta_i - \delta_j - \theta_{Lij}) \nonumber  \\
&- \frac{1}{2}v_i^2 Y_{Sij} \sin(\theta_{Sij}),
\qquad \quad  \qquad  \forall (i,j) \in \mathcal{L} 
\label{eq.5}
\end{IEEEeqnarray}

The active and reactive power lower and upper bound  of each  generation units are defined as: 
\begin{IEEEeqnarray}{l"l}
p_i^{\text{G},\text{min}} \leq p_i \leq p_i^{\text{G},\text{max}}, &  \forall i \in \mathcal{N}   \label{eq.6} \\
q_i^{\text{G},\text{min}} \leq q_i  \leq q_i^{\text{G},\text{max}}, & \forall i \in \mathcal{N}   \label{eq.7} 
 \end{IEEEeqnarray}

\par The line capacity, nodal voltage limits, and capacitor banks constraints are formulated in \eqref{eq.8}, \eqref{eq.9} and \eqref{eq.10}, respectively: 
\begin{IEEEeqnarray}{l"l}
 p_{ij}^2 + q_{ij}^2  \leq (s_{ij}^{\text{max}})^2,    &  \forall (i,j) \in \mathcal{L}  \label{eq.8}  \\
 v_i^{\text{min}} \leq v_i \leq v_i^{\text{max}}, &  \forall i \in \mathcal{N}   \label{eq.9} \\
 0 \leq q_i^{\text{C}}\leq q_i^{\text{C},\text{max}}, &   \forall i \in  \mathcal{N}  \label{eq.10} 
 \end{IEEEeqnarray}

\par Finally, the boundary conditions at the substation are given by: 
\vspace{-0.5cm}

\begin{IEEEeqnarray}{l'l}
 v_1 = 1,   \label{v_ini}\\
   p^{\text{se}} = p_1^{\text{G}}, & q^{\text{se}} = q_1^{\text{G}}  \label{vslack} 
 \label{pslack}
\end{IEEEeqnarray}

\subsection{\distflow Formulation}

\begin{figure}[t]
\centering
    \includegraphics[width=1\linewidth]{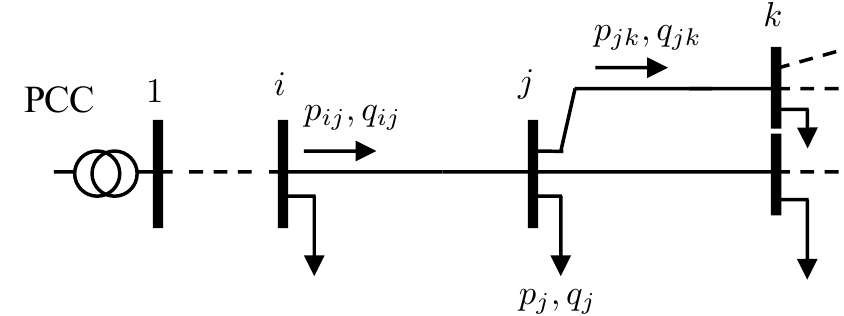}
    \caption{Power flow notations in a radial network}
    \label{fig.scheme}
\end{figure}

The \OPF for distribution network could be represented by the single-phase recursive branch-flow equations based on the Kirchhoff's and the Ohm's law at every bus; known as \distflow equations. This formulation is introduced in the seminal works of \textit{Baran and Wu} \cite{baran1989optimal}, \cite{baran1989optimalPlacement}.  Fig. \ref{fig.scheme} provides notation used in here for power flow equation in radial distribution networks. Thus, the \distflow equations are formulated as:
\vspace{-0.2cm}
\begin{IEEEeqnarray}{l'l}
 p_{ij} = p_{j} + r_{ij}l_{ij}+\sum_{k:(j,k)\in \mathcal{L}} p_{jk},&  \forall (i,j) \in \mathcal{L} \label{eq.1 distflow}\\
q_{ij} = q_{j} + x_{ij}l_{ij}+\sum_{k:(j,k)\in \mathcal{L}}q_{jk},&  \forall (i,j) \in \mathcal{L}  \label{eq.2 distflow}
\\
v_j^2 = v_i^2  + (r_{ij}^2 + x_{ij}^2)l_{ij} \nonumber \\
\qquad \qquad  \qquad   - 2 (r_{ij} p_{ij} + x_{ij} q_{ij}),&  \forall (i,j) \in \mathcal{L}  \label{eq.3 distflow} \\
 p_{ij}^2 + q_{ij}^2 = l_{ij} v_i^2 ,&  \forall (i,j) \in \mathcal{L}  \label{eq. 4 distflow} 
\end{IEEEeqnarray}

\noindent  where $p_{ij}$ and $q_{ij}$ are active and reactive branch power flow from node $i$ to $j$. $l_{ij}$, $r_{ij}$ and $x_{ij}$ are the current squared, resistance, and reactance of the branch $ij$, respectively. $p_{j}$ and $q_{j}$ are the net active and reactive power withdraw at node $j$, represented by \eqref{eq.5 distflow} and \eqref{eq.6 distflow}, respectively.
\begin{IEEEeqnarray}{l'l}
 p_{j} =  p_j^{\text{D}} - p_j^{\text{G}},  & \forall j \in \mathcal{N} \label{eq.5 distflow} \\ 
 q_{j} =  q_j^{\text{D}} -  q_j^{\text{C}} - q_j^{\text{G}},  & \forall j \in \mathcal{N}  \label{eq.6 distflow}
\end{IEEEeqnarray}

Note that power flow equations \eqref{eq2}--\eqref{eq.5} from \acopf formulation are replaced by the equations \eqref{eq.1 distflow}--\eqref{eq.6 distflow} in the \distflow formulation. The later, has  linear definition for the branch flows, \eqref{eq.1 distflow} and \eqref{eq.2 distflow}, while the \acopf are non-linear. Yet, \eqref{eq.2 distflow} and \eqref{eq.3 distflow} are non-linear and non-convex. The technical constraints concerning voltage, generation, and line limits are the same for the \distflow. 

\subsection{\distflowSOCP Formulation}

In this subsection we  transform the non-convex \distflow formulation into a convex one. First, we replace \mbox{$v_i^2 = w_i$}. Thus, the non-linear equation \eqref{eq.3 distflow} is exactly reformulated as a linear one, \eqref{eq.3 distflowSOCP}. 
\begin{IEEEeqnarray}{ll}
 w_j = w_i  + (r_{ij}^2 + x_{ij}^2)l_{ij}   - 2 (r_{ij} p_{ij} + x_{ij} q_{ij}), &  \: \forall (i,j) \in \mathcal{L}   \qquad \label{eq.3 distflowSOCP} 
\end{IEEEeqnarray}

\begin{figure}[t]
\center\includegraphics[width=0.8\linewidth]{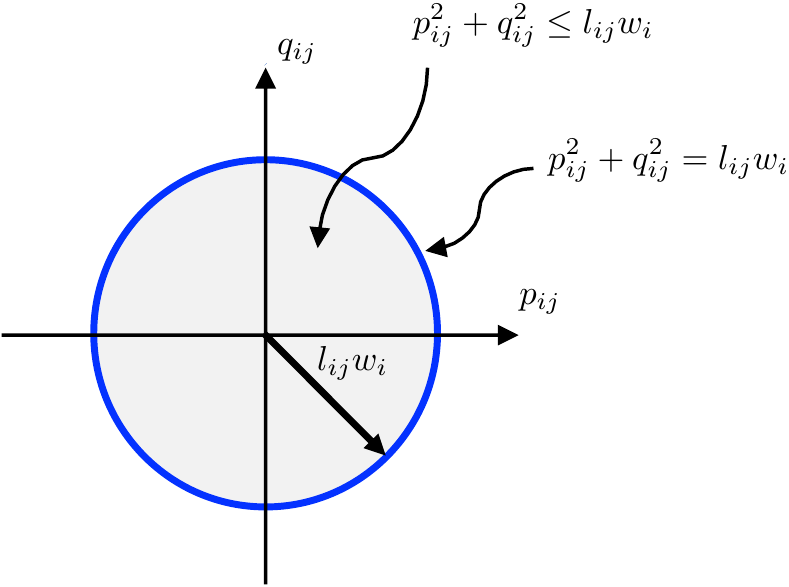}
\caption{Equation \eqref{eq. 4 distflow} and its relaxation \eqref{eq. rel_1} projected onto $p_{ij}$--$q_{ij}$ orthant}
\label{fig.relaxation}
\end{figure}

Next, we can relax the the non-convex equality \eqref{eq. 4 distflow} by a convex inequality constraint as follows (see Fig. \ref{fig.relaxation} for graphical interpretation):
\begin{IEEEeqnarray}{l"l}
 p_{ij}^2 + q_{ij}^2 \leq  l_{ij}w_i,  & \forall (i,j) \in \mathcal{L}  \quad  \label{eq. rel_1}
\end{IEEEeqnarray}

\noindent where term $l_{ij}w_i$ is still a bilinear term, but it can be reformulated as  $l_{ij}w_i =  (\frac{l_{ij} + w_i}{2})^2 - (\frac{l_{ij} - w_i}{2})^2$. Therefore, equation \eqref{eq. rel_1} can be recast as the convex conic constraint \eqref{eq. rel_2}. 
\begin{IEEEeqnarray}{ll}
 p_{ij}^2 + q_{ij}^2 +  \left(\frac{l_{ij} - w_i}{2}\right)^2 \leq 
 \left(\frac{l_{ij} + w_i}{2} \right)^2,  & \;  \forall (i,j) \in \mathcal{L}   \quad \label{eq. rel_2}
\end{IEEEeqnarray}

The resulting \OPF problem is convex; in particular, this problem is classified as a second-order cone programming (\SOCP). We refer to this \OPF formulation as \distflowSOCP.  Existing methods for solving this class of problems guarantee global solutions \cite{alizadeh2003second}. 

As in the \distflow formulation, technical constraints for the system operation are the same as in the \acopf formulation. However, in this case, we need to reformulate constraints related to the voltage magnitude. Therefore, voltage limits \eqref{eq.9} are replaced by \eqref{eq.9 new}. Likewise, boundary condition \eqref{v_ini} at substation is substituted by \eqref{w_ini}.
\begin{IEEEeqnarray}{l"l}
 (v_i^{\text{min}})^2 \leq w_i \leq (v_i^{\text{max}})^2, &  \forall i \in \mathcal{N}   \label{eq.9 new} \\
w_1 = 1  \label{w_ini}
\end{IEEEeqnarray}

\subsection{\lindistflow Formulation}
\label{eq. ADN_OPF}

\textit{Baran and Wu} \cite{baran1989optimal} introduced a linearized version of the \distflow, the \lindistflow model. Later, \textit{Low} reviewed it in \cite{low2014convex}. \lindistflow assumes no losses. The \lindistflow is reformulated from the extract \OPF formulation for radial networks (\distflow formulation), resulting in a linear approximation, attractive for engineers and economists many reasons. The \lindistflow power flow equations  are stated as follow:
\begin{IEEEeqnarray}{l'l}
 p_{ij} = p_{j} + \sum_{k:(j,k) \in \mathcal{L}}p_{jk}, & \forall (i,j) \in \mathcal{L}   \label{lin1}  \\
 q_{ij} = q_{j}+\sum_{k:(j,k)\in \mathcal{L}}q_{jk}, & \forall (i,j) \in \mathcal{L}   \label{lin2}\\
 w_i - w_j =  2(r_{ij} p_{ij} + x_{ij} q_{ij}), &  \: \forall (i,j) \in \mathcal{L}.   \qquad  \label{lin3}
\end{IEEEeqnarray}

The technical constraints and boundary conditions are set like in the \acopf formulation, except for line limits constraints \eqref{eq.8}, that is substituted by \eqref{eq.8 new1} in order to linearize have a full linear model.  
\begin{IEEEeqnarray}{l'l}
p_{ij} \leq p_{ij}^{\text{max}}, q_{ij} \leq q_{ij}^{\text{max}},   &  \forall (i,j) \in \mathcal{L}   \label{eq.8 new1}  
\end{IEEEeqnarray}

\subsection{\OPF Formulation Summary}

In Table \ref{tab. opf models} we have summarized the set of constraints and objective function for each of the fourth formulations that we would like to analyze for flexibility region construction. The main differences among formulations are in the power flow representation. While \acopf and \distflow contains exact representation of the power flows equations, \distflowSOCP is relaxation, and \lindistflow is an approximation of the true power flow equations.

\begin{table*}[ht!]
	\centering
	\caption{Summary for the single-period \OPF models used for feasible region construction}
	\label{tab. opf models}
	\resizebox{\textwidth}{!}{    \begin{tabular}{|r||c|c|c|c|}
			\hline
			Set of constraints/objective & \textbf{\acopf} & \textbf{\distflow} & \textbf{\distflowSOCP} & \textbf{\lindistflow}  \\ 
			\hline
			\textbf{Objective} & \eqref{eq.obFunACOPF}  & \eqref{eq.obFunACOPF}  & \eqref{eq.obFunACOPF} &\eqref{eq.obFunACOPF}  \\
			\hline
			\textbf{Power Flow Constraints} & \eqref{eq2}--\eqref{eq.5} &  \eqref{eq.1 distflow}--\eqref{eq.6 distflow} &  \eqref{eq.1 distflow}, \eqref{eq.2 distflow}, \eqref{eq.3 distflowSOCP}, \eqref{eq.9 new}, \eqref{eq.5 distflow}, \eqref{eq.6 distflow} & \eqref{lin1}--\eqref{lin3}, \eqref{eq.5 distflow}, \eqref{eq.6 distflow}\\
			\hline
			\textbf{Technical limits} & 
			\eqref{eq.6}--\eqref{eq.10} & \eqref{eq.6}--\eqref{eq.10} & 
			\eqref{eq.6}--\eqref{eq.8}, \eqref{eq.10},  \eqref{eq.9 new} & \eqref{eq.6}--\eqref{eq.7}, \eqref{eq.10},  \eqref{eq.9 new},\eqref{eq.8 new1}  \\
			\hline
			\textbf{Boundary conditions at SE} & 
			\eqref{v_ini}, \eqref{vslack} & \eqref{v_ini}, \eqref{vslack} & \eqref{vslack},  \eqref{w_ini} & \eqref{vslack},  \eqref{w_ini}  \\
			\hline \hline
			& \textbf{\acopf} & \textbf{\distflow} & \textbf{\distflowSOCP} & \textbf{\lindistflow} \\
			\hline
			\textbf{Type of power flow approach} &  true representation  & true representation & relaxation & approximation \\
			\textbf{Optimization class} & NLP & NLP & SOCP & LP \\
			\hline
		\end{tabular}
	}
\end{table*}

\section{Epsilon-Constraint Method}
\label{section4}

To define a feasible region between \TSODSO, many scenario-based methods, such as Monte Carlo \cite{ageeva2019analysis}, are introduced in the literature.
Scenario-based Monte Carlo simulation provides various operating points as scenarios; the model is a cloud of the operating points, and also the finite number of scenarios treats properly for convex problems. However, it requires an extensive data set, as the large vectors of uncertainty increase the number of scenarios required to construct the region. 
In this paper, we find the operational limits of the \ADN, which implicitly depends on the optimal operating dispatch. 
For doing so, we solve the multi-objective optimization problem for finding the maximum and minimum of active and reactive power at the PCC \eqref{eq.MO}, while satisfying grid feasibility (i.e., power flow constraints, technical limits and boundary conditions -- see Table \ref{tab. opf models} rows 2--4) for given optimal injections.  Grid feasibility is represented by $\mathcal{F}$ in the multi-objective problem \eqref{eq.MO}.

\begin{IEEEeqnarray}{r'll}
\max \text{/} \min \:  & \left\{p^{se},q^{se}\right\}  \label{eq.MO}\\
\text{s.t:} & (p^{se},q^{se}) \in \mathcal{F}(p_i^{*},q_i^{*}, \: i \in \mathcal{N}  \backslash \{1\} ) \nonumber 
\end{IEEEeqnarray}

The solution of \eqref{eq.MO} is a Pareto front (the feasible region in our case) rather than a single operational point. One of the most well-known methods to solve \eqref{eq.MO} is the so-called epsilon-constraint method \cite{yv1971bicriterion}. In this method a primary objective function is considered while the other objective functions are relaxed with the set of constraints. Perturbations on $\epsilon_i$ size are added to the objectives in the set of constraints and solved sequentially.

\begin{figure}[h]
\center\includegraphics[width=1\linewidth]{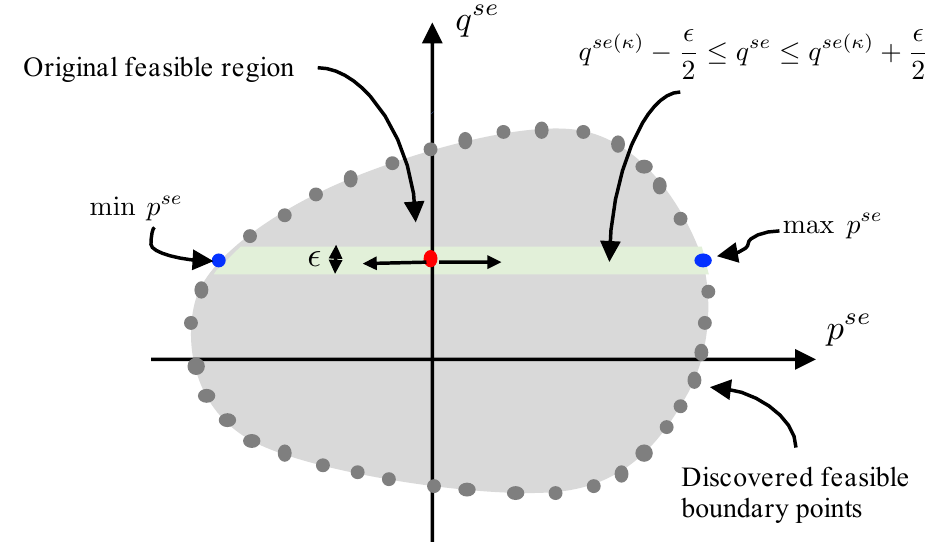}
\caption{Concept of the feasibility region construction by epsilon-constraint method}
\label{fig.LoadProfile}
\end{figure}

Thus, the \TSODSO feasible region can be reconstructed by solving a sequential set of problems. For the particular iteration $\kappa$, the problem to solve is depicted in \eqref{eq.MO_iter}, where $q^{se(\kappa)}$ is updated in each iteration.  Figure \ref{fig.LoadProfile} represents the schematic feasibility region created by the epsilon-constraint method.

\begin{IEEEeqnarray}{r'll}
\max / \min \:  & p^{se} \label{eq.MO_iter}\\
\text{s.t:} & (p^{se},q^{se}) \in \mathcal{F}(p_i^{*},q_i^{*}, \: i \in \mathcal{N}  \backslash \{1\} ) \nonumber \\
& q^{se(\kappa)} - \frac{\epsilon}{2} \leq q^{se} \leq q^{se(\kappa)} + \frac{\epsilon}{2} \nonumber
\end{IEEEeqnarray}

The stopping criterion is set for a given number $N$ of points that will form the boundary accordingly to the resolution needed. Then, the value of $\epsilon$ can be computed to equally divide the continuous space of feasible values of $q^{se}$ \cite{yv1971bicriterion}.

\section{Case Study}
\label{section5}

\begin{figure}[h]
\center
\includegraphics[width=0.9\linewidth]{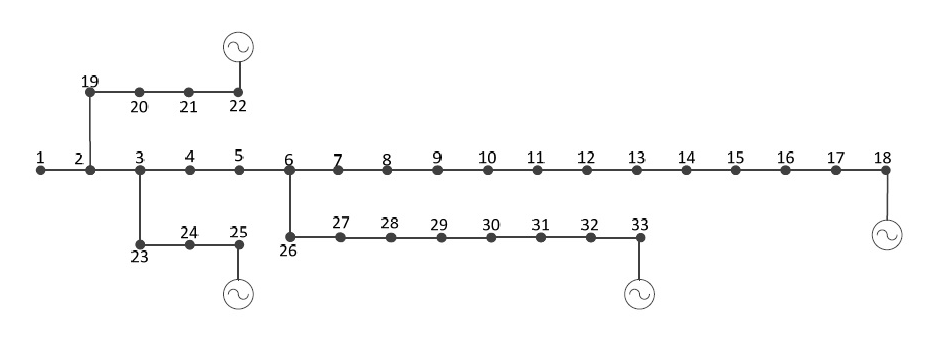}
\caption{33-bus radial distribution system}
\label{fig.33 bus graph}
\end{figure}

We have used the IEEE 33-bus radial distribution system for testing our method. We have modified this case by including four fast distributed generators (DGs) in Fig.  \ref{fig.33 bus graph}.  Capacitor banks of $1$MVAr are connected to nodes $10$, $20$ and $30$.
Node 1 is the slack node and the defines as the PCC, where is the interface between transmission and distribution network. Twenty-four hours were simulated. We chose 200 points for building feasible boundaries for all case studies.

The implementation of the \OPF models was done using Julia 1.4, and JuMP 0.19, with the solvers Ipopt 0.5.4 for NLP problems and Gurobi 9.0. for SOCP and LP problems.  All simulations were performed in a Laptop Intel(R) Core(TM) i5-6200U CPU with installed RAM 8GB. Note that for solving the sequence of NLP in the \acopf and \distflow, we have provided starting point to Ipopt based on previous iteration.

\vspace{0.3cm}

Table \ref{tab:time_Epsilon} summarizes the computation time for each model. We observe the efficiency of \distflow with respect to the \acopf formulation with a time reduction of about $1/3$. The \lindistflow method, of course, is the fastest one, about $780\times$ faster than \acopf, and $290\times$ faster than \distflow.

\begin{table}[h]
 \caption{CPU time for building feasible regions with epsilon \\ constraint method}
 \centering
\begin{tabular}{r| l}
    {\textbf{Model}} & \textbf{CPU time [seconds]} \\
\hline
    {\acopf}     &  1798.3\\
   {\distflow}     &  666.2\\
   \distflowSOCP     & 478.8 \\
    \lindistflow & 2.3  \\
    \hline
    \end{tabular}%
 \label{tab:time_Epsilon}
\end{table}%

\begin{figure}[H]
    \centering
    {\includegraphics[width=1\linewidth]{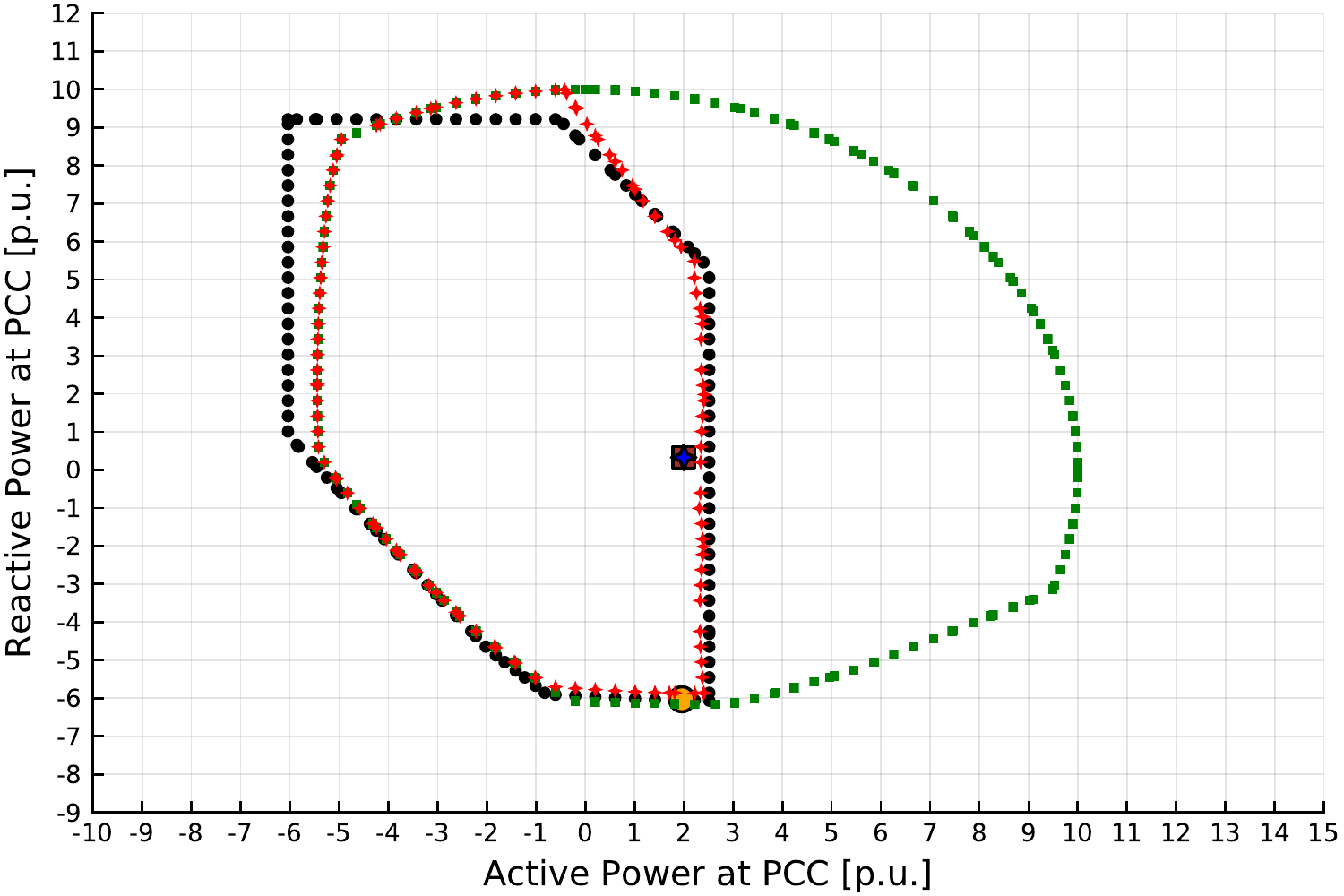}} 
    \caption{Feasibility region constructed with the epsilon constraint method for period 14. Black circles for \lindistflow model, red stars for \distflow and \acopf model, and green squares for \distflowSOCP model.} \label{fig.feasible}
\end{figure}

Figure \ref{fig.feasible} represents the reconstruction for the feasible domain for the hour 14 at the interface \TSODSO. Both \acopf and \distflow are the same and depicted in red. The black circles denote the \lindistflow, and green squares represent the feasible domain when using a relaxed formulation \distflowSOCP.  The relaxed formulation overestimates the active power that the \ADN can absorb (positive orthant). At the same time, the \distflowSOCP quite well approximates the power available to inject into the transmission grid. On the other hand, the \lindistflow is approaching very close to the exact feasible domain. But, it has the main discrepancies in the negative orthant, i.e., active power that could be injected from the \ADN to the transmission grid. This is mainly because the \lindistflow does not consider losses overestimating the capability of energy that could be generated from the \ADN. 

\begin{figure}[h]
\center\includegraphics[width=1\linewidth]{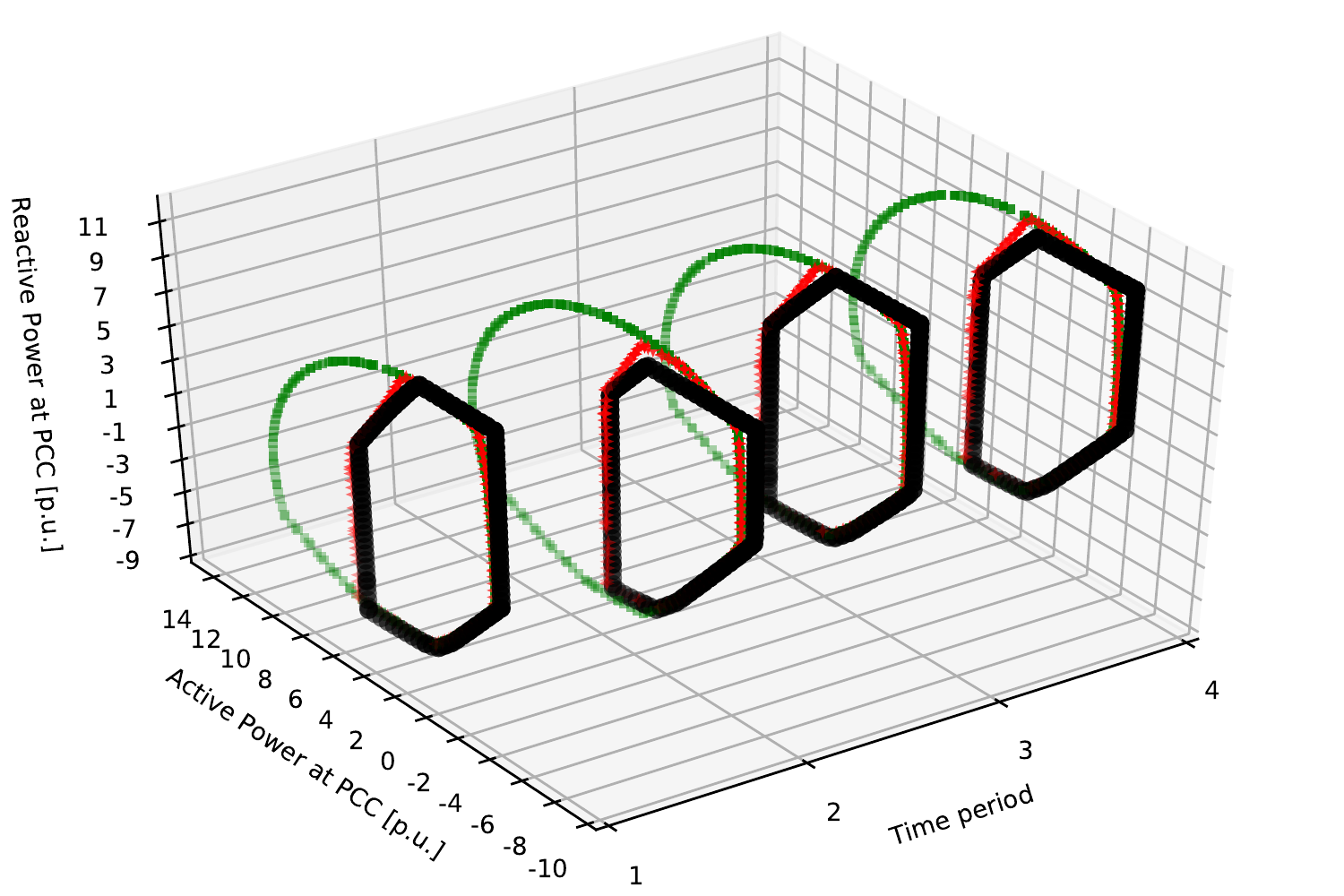}
\caption{Feasibility region for four consecutive periods}
\label{fig.Multiperiod}
\end{figure}

Fig. \ref{fig.Multiperiod} shows four consecutive hours resulting from the feasible region generation. It is worth noting that even we have omitted in our formulation the multi-period \OPF, it has implemented in a multi-period fashion, and it can be easily extended and include inter-temporal constraints related to energy storage or ramping capabilities. 

\section{Conclusions and Discussion}
\label{section6}

In this paper, we have defined the flexibility region of an \ADN (\DSO level) connected to a \TSO based on mathematical programming and power systems principles.  This work evaluates and compares different optimal power flow formulations (\acopf, \distflow, \distflowSOCP, and \lindistflow), commonly used in the literature, for building feasible regions for \TSODSO interaction.   
An epsilon-constraint method is proposed for constructing the feasible operation region at the \TSODSO interface.    

The presented method is analyzed in an IEEE 33-bus test radial distribution system. The solutions obtained are compared for each model and proposed. 
We have observed that \lindistflow is fast and relatively closed to the exact original feasible region. On the other hand, the convex relaxation \distflowSOCP showed large discrepancies with regard to the exact feasible limits. This raises an attention note on the use of convexification models. 
Finally, results from this research work could help to contribute in future and open challenges for an advanced \TSODSO cooperation.

\section{Acknowledgements}
This work was supported by Skoltech NGP Program (Skoltech-MIT joint project)

\section*{References}

\bibliographystyle{IEEEtran}
\bibliography{ref_mod}

\begin{thebibliography}{10}
\providecommand{\url}[1]{#1}
\csname url@samestyle\endcsname
\providecommand{\newblock}{\relax}
\providecommand{\bibinfo}[2]{#2}
\providecommand{\BIBentrySTDinterwordspacing}{\spaceskip=0pt\relax}
\providecommand{\BIBentryALTinterwordstretchfactor}{4}
\providecommand{\BIBentryALTinterwordspacing}{\spaceskip=\fontdimen2\font plus
\BIBentryALTinterwordstretchfactor\fontdimen3\font minus
  \fontdimen4\font\relax}
\providecommand{\BIBforeignlanguage}[2]{{%
\expandafter\ifx\csname l@#1\endcsname\relax
\typeout{** WARNING: IEEEtran.bst: No hyphenation pattern has been}%
\typeout{** loaded for the language `#1'. Using the pattern for}%
\typeout{** the default language instead.}%
\else
\language=\csname l@#1\endcsname
\fi
#2}}
\providecommand{\BIBdecl}{\relax}
\BIBdecl

\bibitem{barenfanger2016classifying}
R.~B{\"a}renf{\"a}nger, E.~Drayer, D.~Daniluk, B.~Otto, E.~Vanet, R.~Caire,
  T.~S. Abbas, and B.~Lisanti, ``Classifying flexibility types in smart
  electric distribution grids: a taxonomy,'' in \emph{CIRED Workshop
  2016}.\hskip 1em plus 0.5em minus 0.4em\relax IET, 2016, pp. 1--4.

\bibitem{villar2018flexibility}
J.~Villar, R.~Bessa, and M.~Matos, ``Flexibility products and markets:
  Literature review,'' \emph{Electric Power Systems Research}, vol. 154, pp.
  329--340, 2018.

\bibitem{gonzalez2018determination}
D.~M. Gonzalez, J.~Hachenberger, J.~Hinker, F.~Rewald, U.~H{\"a}ger,
  C.~Rehtanz, and J.~Myrzik, ``Determination of the time-dependent flexibility
  of active distribution networks to control their tso-dso interconnection
  power flow,'' in \emph{2018 Power Systems Computation Conference
  (PSCC)}.\hskip 1em plus 0.5em minus 0.4em\relax IEEE, 2018, pp. 1--8.

\bibitem{contreras2018improved}
D.~A. Contreras and K.~Rudion, ``Improved assessment of the flexibility range
  of distribution grids using linear optimization,'' in \emph{2018 Power
  Systems Computation Conference (PSCC)}.\hskip 1em plus 0.5em minus
  0.4em\relax IEEE, 2018, pp. 1--7.

\bibitem{ji2018robust}
T.~Ji, J.~Yao, Y.~Wu, A.~Ehsan, M.~Cheng, and Q.~Yang, ``Robust active
  distribution network planning considering stochastic renewable distributed
  generation,'' in \emph{2018 37th Chinese Control Conference (CCC)}.\hskip 1em
  plus 0.5em minus 0.4em\relax IEEE, 2018, pp. 8803--8808.

\bibitem{elliott2018sharing}
R.~T. Elliott, R.~Fernandez-Blanco, K.~Kozdras, J.~Kaplan, B.~Lockyear,
  J.~Zyskowski, and D.~S. Kirschen, ``Sharing energy storage between
  transmission and distribution,'' \emph{IEEE Transactions on Power Systems},
  vol.~34, no.~1, pp. 152--162, 2018.

\bibitem{ageeva2019analysis}
L.~Ageeva, M.~Majidi, and D.~Pozo, ``Analysis of feasibility region of active
  distribution networks,'' in \emph{2019 International Youth Conference on
  Radio Electronics, Electrical and Power Engineering (REEPE)}.\hskip 1em plus
  0.5em minus 0.4em\relax IEEE, 2019, pp. 1--5.

\bibitem{silva2018estimating}
J.~Silva, J.~Sumaili, R.~J. Bessa, L.~Seca, M.~A. Matos, V.~Miranda,
  M.~Caujolle, B.~Goncer, and M.~Sebastian-Viana, ``Estimating the active and
  reactive power flexibility area at the tso-dso interface,'' \emph{IEEE
  Transactions on Power Systems}, vol.~33, no.~5, pp. 4741--4750, 2018.

\bibitem{pastor2018evaluation}
A.~V. Pastor, J.~N. Martin, D.~W. Bunn, and A.~Laur, ``Evaluation of
  flexibility markets for retailer-dso-tso coordination,'' \emph{IEEE
  Transactions on Power Systems}, 2018.

\bibitem{wood2013power}
A.~J. Wood, B.~F. Wollenberg, and G.~B. Shebl{\'e}, \emph{Power generation,
  operation, and control}.\hskip 1em plus 0.5em minus 0.4em\relax John Wiley \&
  Sons, 2013.

\bibitem{molzahn2019survey}
D.~K. Molzahn, I.~A. Hiskens \emph{et~al.}, ``A survey of relaxations and
  approximations of the power flow equations,'' \emph{Foundations and
  Trends{\textregistered} in Electric Energy Systems}, vol.~4, no. 1-2, pp.
  1--221, 2019.

\bibitem{baran1989optimal}
M.~Baran and F.~F. Wu, ``Optimal sizing of capacitors placed on a radial
  distribution system,'' \emph{IEEE Transactions on power Delivery}, vol.~4,
  no.~1, pp. 735--743, 1989.

\bibitem{baran1989optimalPlacement}
M.~E. Baran and F.~F. Wu, ``Optimal capacitor placement on radial distribution
  systems,'' \emph{IEEE Transactions on power Delivery}, vol.~4, no.~1, pp.
  725--734, 1989.

\bibitem{alizadeh2003second}
F.~Alizadeh and D.~Goldfarb, ``Second-order cone programming,''
  \emph{Mathematical programming}, vol.~95, no.~1, pp. 3--51, 2003.

\bibitem{low2014convex}
S.~H. Low, ``Convex relaxation of optimal power flow—part {I}: Formulations
  and equivalence,'' \emph{IEEE Transactions on Control of Network Systems},
  vol.~1, no.~1, pp. 15--27, 2014.

\bibitem{yv1971bicriterion}
Y.~Haimes, L.~S. Lasdon, and D.~Wismer, ``On a bicriterion formation of the
  problems of integrated system identification and system optimization,''
  \emph{IEEE Transactions on Systems, Man and Cybernetics}, no.~3, pp.
  296--297, 1971.

\end{thebibliography}

\end{document}